\documentclass[
    ,final            
  ]
  {aipproc}

\usepackage{natbib}
\usepackage{amssymb}
\layoutstyle{8x11single}

\begin{document}

\title{The innocuousness of adiabatic instabilities in coupled
  scalar field-dark matter models}


\author{P.S. Corasaniti}{
  address={LUTH, Observatoire de Paris, CNRS UMR 8102, Universit\'e
  Paris Diderot, 5 Place Jules Janssen, 92195 Meudon Cedex, France}
}

\begin{abstract}
Non-minimally coupled scalar field models suffer of unstable
growing modes at the linear perturbation level. The nature of these
instabilities depends on the dynamical state of the scalar field. 
In particular in systems which admit adiabatic solutions, 
large scale instabilities are suppressed by the slow-roll 
dynamics of the field. Here we review these results and present 
a preliminary likelihood data analysis suggesting
that along adiabatic solutions coupled models with coupling of order
of gravitational strength can provide viable cosmological scenarios satisfying
constraints from SN Ia, CMB and large scale structure data.
\end{abstract}

\maketitle


\section{Introduction}
The possibility of a direct coupling between a quintessence-like
scalar field and the various matter components has been extensively studied in a vast literature
(see e.g. \cite{LucaDomenico}). Non-minimally coupled scalars appear in various theoretical 
scenarios which attempt to describe fundamental interactions at energies 
beyond that of the Standard Model of particle physics (see e.g. \cite{Damour}). Their 
application to cosmology and the unsolved problem of dark energy
in the universe has suggested a number of interesting features, 
most importantly the solution of the so
called coincidence problem. The presence of the scalar interaction is
not incompatible with existing constraints on the violation of the
Equivalence Principle. Ingenuous mechanisms (which may differ from one model to another) guarantee
that standard General Relativity is recovered on Solar System scales
(e.g. \cite{Justin,Alimi}), and leave 
distinctive signatures on the structure formation process, eventually
contributing to some of the still not understood phenomena in the
context of Cold Dark Matter paradigm \cite{PeeblesFarrar}. 
Consequently the distribution of structures, at
least on those scales where observations have provided accurate
measurements, is a key test that such models have to pass. 
Nevertheless a number of works have indicated that coupled 
scalar field models may suffer of large scale instabilities at the
linear perturbation level. This was initially pointed out in some
specific realizations \cite{KoivistoKaplingat} and recently
discussed in more general setups \cite{Bean,ValiviitaAbdalla}. The
claim has been particularly emphasized on scenarios characterized by
the existence of the so called ``adiabatic'' regime, such as the
Chameleon model \cite{Bean}. In the light of these results coupled models
seem to be unrealistic cosmological scenarios. However as we have shown in
\cite{PSC} the instabilities are not generic, rather they are
strongly dependent on the dynamical state of the
scalar field. More importantly  along ``adiabatic'' solutions
and for natural values of the coupling constant such instabilities are innocuous. 
Here we will briefly summarize the main results of
\cite{PSC} to which we refer the reader for a more detailed
discussion. We will also present the results of a preliminary
likelihood data analysis to test the viability of these models against
current cosmological observations. 
 
\section{Perturbations in coupled scalar field-dark matter models}

Let us consider a scalar field $\phi$, with potential $V(\phi)$, coupled to 
matter particles through a Yukawa coupling of the form $f(\phi/M_{Pl})\psi\bar{\psi}$,
where $f$ is the coupling function and $\psi$ is the Dirac spinor associated with the
matter particle ($M_{Pl}=1/\sqrt{8\pi G}$ with G the Newton constant).
For simplicity let us consider the case in which the scalar field is
coupled to dark matter only, thus Equivalence Principle constraints are immediately satisfied.
This is not a restrictive assumption since our results can be extended also to models with couplings to all matter
components provided the existence of an ``adiabatic'' regime. 
Because of the coupling the energy-momentum tensor of each component
of the system is not conserved. It is
only the total energy-momentum tensor that satisfies the conservation equation: $T_{\nu;\mu}^{\mu(T)}\equiv T_{\nu;\mu}^{\mu(\phi)}+T_{\nu;\mu}^{\mu(DM)}=0$.
Now let us consider a coupling function of dilatonic type, $f(\phi)=exp(\beta\phi/M_{Pl})$,
with $\beta$ the dimensionless coupling constant, from the above
conservation condition in a flat Friedmann-Lemaitre-Robertson-Walker background ($ds^2=-dt^2+a(t)^2d\textbf{x}^2$)
we obtain the evolution equations:
\begin{eqnarray}
\dot{\rho}_{DM}+3H\rho_{DM}&=&\beta \dot{\phi} \rho_{DM},\label{rhom}\\ 
\ddot{\phi}+3H\dot{\phi}+V_{,\phi}&=&-\beta\rho_{DM},\label{KG}
\end{eqnarray}
 with the Hubble rate given by $H^2\equiv\left(\frac{\dot{a}}{a}\right)^2=\frac{1}{3}\left[\rho_{DM}+\dot{\phi}^2/2+V(\phi)\right]$.
The solution to Eq.~(\ref{rhom}) reads as
\begin{equation}
\rho_{DM}=\frac{\rho_{DM}^{(0)}}{a^3}e^{\beta(\phi-\phi_0)},
\end{equation}
where $\phi_0$ is the present scalar field value. From Eq.~(\ref{rhom})
and Eq.~(\ref{KG}) we may notice that for positive values of the coupling constant $\beta$, the interaction transfers
energy from the $\phi$-field to the dark matter particles, with the scalar field evolving in an
effective potential 
\begin{equation}
V_{\rm eff}(\phi)=V(\phi)+\frac{\rho_{DM}^{(0)}}{a^3}e^{\beta(\phi-\phi_0)},
\end{equation}
characterized by a minimum 
\begin{equation}
V_{,\phi_{\rm min}}=-\beta
\frac{\rho^{(0)}_{DM}}{a^3}e^{\beta(\phi_{\rm min}-\phi_0)}.\label{adiacond}
\end{equation}

Given the above background equations, the evolution of linear density fluctuations can be studied by perturbing
the Einstein equations and the conservation of the total energy momentum tensor
about a linearly perturbed FLRW background. However in order to gain
some intuitive insight on the behavior
of the perturbations on the large scales and perform a simple stability analysis, it is convenient to
consider the interacting scalar field-dark matter system as an
effective single fluid with
energy density $\rho_{T}=\dot{\phi}^2/2+V(\phi)+\rho_{DM}$, pressure $p_T=\dot{\phi}^2/2-V(\phi)$,
and whose perturbations are uniquely characterized by an adiabatic sound speed, $c_{aT}=\sqrt{\dot{p}_T/\dot{\rho}_T}$,
and the rest frame sound speed, $c_{sT}=\sqrt{\delta{p_T}/\delta{\rho_t}}$. 

In synchronous gauge the perturbation equations reads as
\begin{eqnarray}
\dot{\delta}_T&=&-3H(c_{sT}^2-w_T)\delta_T+\nonumber\\
&-&(1+w_T)\left\{\left[\frac{k^2}{a^2 H^2}+9(c_{sT}^2-c_{aT}^2)\right]
\frac{a H^2}{k^2} \theta_T+\frac{\dot{h}}{2}\right\},\nonumber \\ \\
\dot{\theta}_T&=&-H(1-3 c_{sT}^2)\theta_T+\frac{c_{sT}^2 k^2}{a(1+w_T)}\delta_T,
\end{eqnarray}
where $\delta_T=\delta{\rho_T}/\rho_T$ and $\theta_T$ is the shear
velocity perturbation of the fluid.  For a barotropic component with a constant
equation of state (e.g. matter, radiation) $c^2_s=c^2_a=w$. 
This is not the case for a generic fluid 
(e.g. scalar field), for this reason we may 
expect the effective unified fluid to be non-barotropic, (i.e.
$c^2_{sT}\neq c^2_{aT} \neq w_T$). 
In terms of the scalar field and dark matter perturbation variables 
we have
\begin{eqnarray}
c^2_{aT}&=&\frac{3H\dot{\phi}^2+\dot{\phi}[2 V_{,\phi}+\beta
    \rho_{DM}]}{3H\dot{\phi}^2+3H\rho_{DM}},\label{ca}\\
c^2_{sT}&=&\frac{\dot{\phi}\delta\dot{\phi}-V_{,\phi}\delta\phi}{\dot{\phi}\delta\dot{\phi}+V_{,\phi}\delta\phi+\rho_{DM}\delta_{DM}}.\label{cs}
\end{eqnarray}
These relations provide us with a simple way of determining the
stability of the perturbations in the coupled system,
for example in a given background regime 
instabilities may develop if these 
sound speeds acquire sufficiently negative values.

\section{Scalar Field Dynamics and Instability Analysis}
An attractor solution of the background homogeneous system consists of the field seating at the
minimum of the potential, and drifting in time according to
Eq.~(\ref{adiacond}). This is usually referred as ``adiabatic'' regime. 
It has been shown in \cite{Das} that along this solution the field slow-rolls, thus it has
a negligible kinetic energy. In particular for a power law potential,
$V(\phi)\propto \phi^{-\alpha}$,
the evolution of the scalar field given by the condition Eq.~(\ref{adiacond}) reads as:
\begin{equation}
\left(\frac{\phi_0}{\phi_{\rm min}}\right)^{\alpha+1}=\frac{1}{a^3}e^{\beta(\phi_{\rm min}-\phi_0)},\label{fieldadia}
\end{equation}
which depends on both the slope $\alpha$ and the coupling
$\beta$. Equation~(\ref{fieldadia}) 
is a non-linear algebraic equation which can be solved numerically through
standard bisection methods.
\begin{figure}[t]
 \includegraphics[scale=0.45]{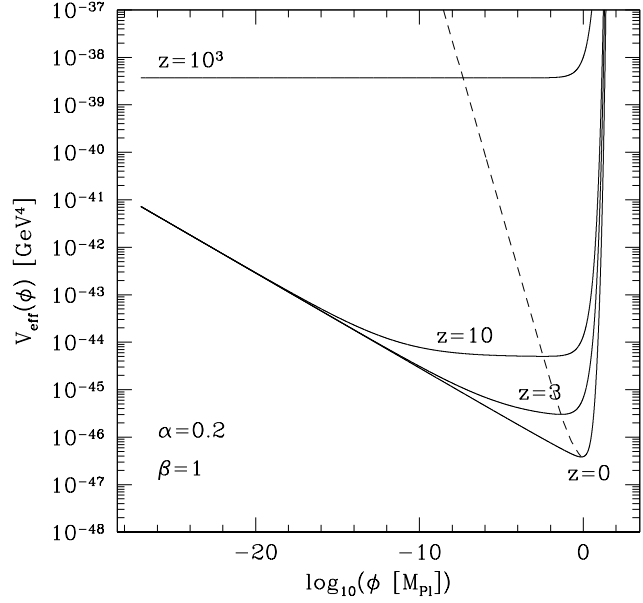} 
  \caption{Scalar field effective potential at $z=0,3,10$ and $10^3$,
  the dashed line shows the position of the minimum as function of the
  redshift.}\label{fig1}
\end{figure}
The presence of the minimum distinguishes two different sets of initial
conditions:
$\phi_{ini}<\phi^{ini}_{\rm min}$ (small field) or $\phi_{ini}>\phi^{ini}_{\rm min}$ (large field).
For small field values, $\phi$ evolves over the inverse
power-law part of the effective potential, where it minimizes 
the potential by slow-rolling as shown in \cite{Das}. 
In contrast for initially large field values, $\phi$ rolls
towards the minimum along the steep exponential part of the effective potential.
Thus it rapidly acquires kinetic energy which subsequently dissipates 
through large high-frequency damped oscillations around the minimum. 
The growth of the linear perturbations in the coupled scalar field-dark
matter system is significantly different in these two regimes.

\subsection{Adiabatic Regime: slow-roll suppresion of instabilities}
\begin{figure}[t]
\includegraphics[scale=0.45]{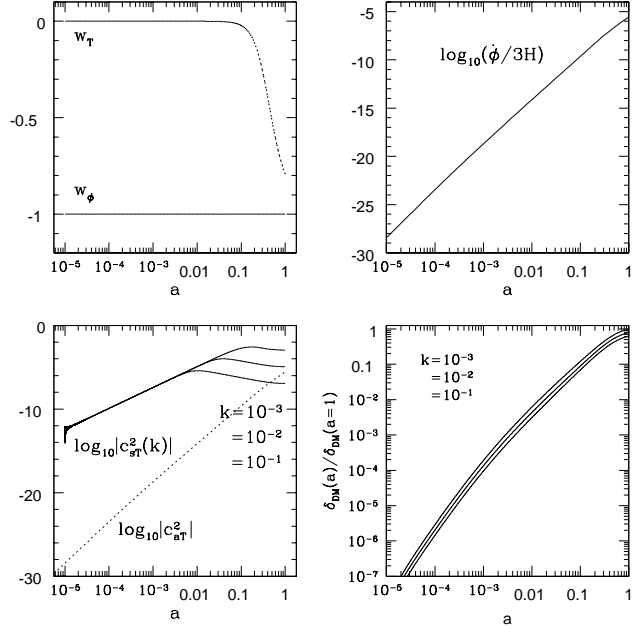}
\caption{Upper left panel: evolution of the scalar field equation of
  state $w_\phi$ and effective unified fluid equation of state $w_T$;
  Right upper panel: evolution of the scalar field velocity with
  respect to the Hubble rate; Lower left panel: redshift evolution of
  the adiabatic sound speed $c_{aT}^2$ and propagation of pressure
perturbations $c_{sT}^2$;
  Right lower panel: Linear growth factor of the dark matter density
  contrast at $k=10^{-3},10^{-2}$ and $0.1$ Mpc$^{-1}$ .}
\label{fig2}
\end{figure}

Let us evaluate the adiabatic and rest frame sound speeds along the adiabatic
solution respectively. Substituting Eq.~(\ref{adiacond}) in
Eq.~(\ref{ca}) and neglecting the term proportional to the kinetic
energy of the scalar field (due to the slow-roll condition) we have
\begin{equation}
c_{aT}^2=-\beta\frac{\dot{\phi}}{3H},\label{cs_adia_cond}
\end{equation}
since $\dot{\phi}>0$ it then follows that $c_{aT}^2<0$, 
implying that adiabatic instabilities may indeed develop. 
However we should remark that during the adiabatic
regime the field is slow-rolling
(i.e. $3H\dot{\phi}\approx 0$), hence 
the term $\dot{\phi}/3H$ can be negligibly small compared 
to $\beta$, such that $c_{aT}^2\approx
0^-$, thus leading to a stable growth of the large scale
perturbations. In fact let us suppose that $c_{aT}^2=-10^{-5}$,
the instability will affect modes $k\ge10^5$, but these
correspond to very small scales which are already in the non-linear
regime and for which the linear perturbation theory does not apply any
longer. 
In contrast large scale instabilities will occur if the coupling assumes extremely large values, 
$\beta\gg 3H/\dot{\phi}$. 
This is consistent with the conclusions of \cite{Bean}, 
where the authors have shown that 
during the adiabatic regime perturbations suffer of instabilities
provided that $\beta\gg 1$. However such situation would be extremely
unnatural introducing a large hierachy problem in the gravitational
sector since it would implying
having a scalar fifth-force which is $(1+2\beta^2)$ greater than
gravitational strength. Guided by naturalness considerations one might expect that the dimensionless coupling
constant is of order of unity. Let us now evaluate the sound
speed in the total effective fluid rest frame, 
Eq.~(\ref{cs}) we have
\begin{equation}
c_{sT}^2=-\frac{1}{1-\frac{1}{\beta}\frac{\delta_{DM}}{\delta\phi}},
\end{equation}
assuming that the scalar field is nearly homogeneous, 
$\delta\phi\ll\delta_{DM}$ (in Planck units), we have
$c_{sT}^2\approx \beta\delta\phi/\delta_{DM}$,
and for $\beta\approx \mathcal{O}(1)$ this implies $c_{sT}^2\approx 0$.
In other words if the scalar field fluctuations are small
with respect to the dark matter density contrast, 
then the coupled system behaves has
a single adiabatic inhomogeneous fluid ($c_{sT}^2\approx
c_{aT}^2\approx 0$).
These results are supported by 
the numerical study of the perturbation equations for the individual
components of the system as summarized in Fig.~\ref{fig2}.

\subsection{Non-Adiabatilsc Regime: large field oscillations and onset of instabilities}
\begin{figure}[t]
\includegraphics[scale=0.45]{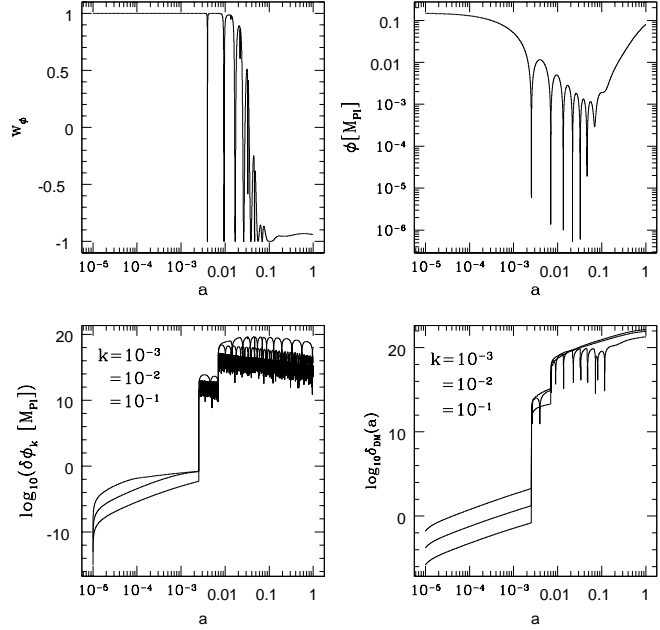}
\caption{Upper left panel: evolution of the scalar field equation of
  state $w_\phi$; Right upper panel: evolution of the scalar field; 
Lower left panel: evolution of the field fluctuations $\delta\phi_{k}$
 at $k=10^{-3},10^{-2}$ and $0.1$ Mpc$^{-1}$ respectively;
  Right lower panel: evolution of dark matter density for $k$-values
  as in the case of $\delta\phi_k$.}
\label{fig3}
\end{figure}

For initially large field values, $\phi$ rolls along the
steep exponential part of the effective potential. Its evolution is
therefore dominated by the kinetic energy and the field behaves as a
stiff fluid ($w_\phi=1$, as can be noticed in the left upper panel of
Fig.~\ref{fig3}). Then as the field reaches the minimum, 
the kinetic energy is damped away through a series of high-frequency
oscillations. During this oscillatory regime, which is similar to that
of the inflaton in the reheating phase, the scalar field perturbations
are unstable and exponentially amplified by the background-field
oscillations provided that their frequency increases as their amplitude
diminishes \cite{kamion}. This is indeed the case as shown in
Fig.~\ref{fig3}, where we plot the evolution of $\phi$ (right upper
panel), $\delta\phi_k$ and $\delta_{DM}$ for three different
wave-numbers, $k=10^{-3},10^{-2}$ and $10^{-1}$.
We can see there that an instability
occurs roughly at the same time of the first minimum-crossing oscillation, 
then followed by a second stage of exponential growth at the beginning 
of the second oscillation.
Such unstable modes are similar to those found in
\cite{ValiviitaAbdalla}, in fact by averaging over periods of time larger 
than the characteristic time of the oscillations, the scalar field 
behaves effectively as a dark energy fluid with a constant equation
of state $w$, as the case considered in \cite{ValiviitaAbdalla}. 

\section{Constraints from SN Ia, CMB and LSS: a preliminary analysis}
\begin{figure}[t]
\includegraphics[scale=0.45]{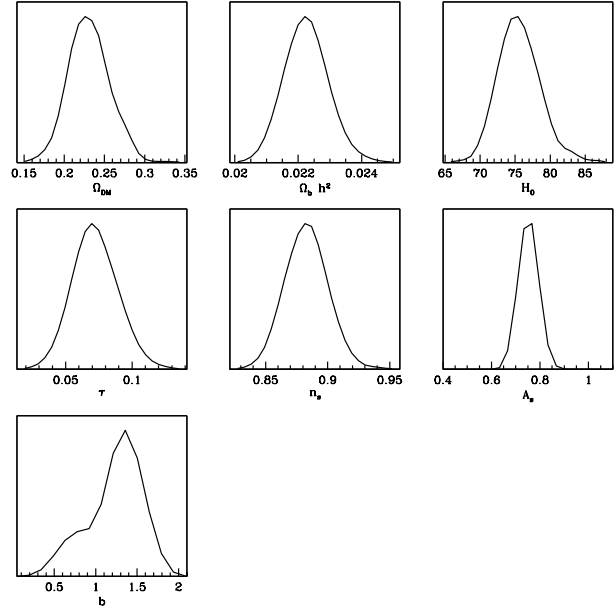}
\caption{Marginalized $1$D likelihoods.}
\label{fig4}
\end{figure}

Coupled models have been tested against cosmological observations in
various works \cite{bean08,mena09,lavacca09}. The main conclusion of these analysis is that
current measurements of the CMB anisotropy power spectra and the matter power
spectrum from galaxy surveys constrain the coupling constant to be
$\beta<0.01-0.1$ depending on the specific model realization. However none of these works have considered the case
in which the scalar field evolves in the adiabatic regime. To this
purpose we used the lastest SN Ia-UNION dataset compilation \cite{union}, WMAP-5
years data \cite{wmap5} and matter power spectrum measurements from SDSS-5 data
release \cite{sdss} to test the viability of 
a non-minimally coupled scalar field model
with power law potential in the adiabatic regime. For this purpose we
have specifically set the field evolution to satisfy Eq.~(\ref{fieldadia}), implemented the
perturbation equations in a properly modified version of the CMBFAST code
\cite{zalda} and run a Markov Chain Monte Carlo likelihood evaluation.
For simplicity we assume a flat universe and fix $\beta=1$ and $\alpha=0.2$, 
while let all other parameters to vary. Here we simply aimed to test
whether an adiabatic solution can provide compatible fit to the data,
and we leave to a future study a more detailed analysis of the full model
parameter space, including $\alpha$ and $\beta$. The marginalized $1$D likelihood are shown in
Fig.~\ref{fig4} and the best fit value and $1\sigma$ errors are: $\Omega_{DM}=0.222\pm0.025$, $\Omega_b
h^2=0.02226\pm0.00068$, $h=0.75\pm0.03$, $\tau=0.072\pm0.017$,
$n_s=0.89\pm0.02$, $A_s=0.75\pm0.03$ and $b=1.44\pm0.49$. These constraints are
consistent with those derived for LCDM cosmologies, the total
$\chi^2$ is close to that of the vanilla LCDM model such the two
scenarios are statistically indistinguishable. As shown in \cite{Das}
differences due to the scalar interaction may arise on the small scale
clustering of dark matter. Overall this preliminary analysis suggests
that adiabatic coupled models with natural coupling are consistent
with current cosmological observations.
 
\bibliographystyle{aipproc}
\bibliography{psc}

\begin{thebibliography}{68}
\expandafter\ifx\csname natexlab\endcsname\relax\def\natexlab#1{#1}\fi
\providecommand{\enquote}[1]{``#1''}
\expandafter\ifx\csname url\endcsname\relax
  \def\url#1{\texttt{#1}}\fi
\expandafter\ifx\csname urlprefix\endcsname\relax\def\urlprefix{URL }\fi
\providecommand{\eprint}[2][]{\url{#2}}

\bibitem{LucaDomenico} L. Amendola, Phys. Rev. D\textbf{62}, 043511
  (2000); D. Tocchini-Valentini and L. Amendola,
  Phys. Rev. D\textbf{65}, 063508 (2002); L.P. Chimento, A.S. Jakubi, D. Pavon and W. Zimdahl, 
Phys. Rev. D\textbf{67}, 083513 (2003); D. Comelli, M. Pietroni and A. Riotto,
  Phys. Lett. B\textbf{571}, 115 (2003); D.B. Kaplan, A.E. Nelson and N. Weiner,
Phys. Rev. Lett. \textbf{93}, 091801 (2004); A. Fuzfa and J.-M. Alimi, Phys. Rev. D\textbf{73},
023520 (2006)

\bibitem{Damour} T. Damour and A.M. Polyakov,
  Nucl. Phys. B\textbf{423}, 532 (1994)

\bibitem{Justin} J. Khoury and A. Weltman, Phys. Rev. D\textbf{69},
  044026 (2004); J. Khoury and A. Weltman, Phys. Rev. Lett.,
  \textbf{93}, 171104 (2004)

\bibitem{Alimi} J.-M. Alimi and A. Fuzfa, JCAP, 0809, 014 (2008)

\bibitem{PeeblesFarrar} G.R. Farrar and P.J.E. Peebles,
  Astrophys. J. \textbf{604}, 1 (2004);  A. Nusser, S.S. Gubser and 
P.J.E. Peebles, Phys. Rev. D\textbf{71}, 083505 (2005)

\bibitem{KoivistoKaplingat} T. Koivisto, Phys. Rev. D\textbf{72},
043516 (2005); M. Kaplinghat and A. Rajaraman,
Phys. Rev. D\textbf{75}, 103504 (2007)

\bibitem{Bean} R. Bean, E.E. Flanagan and M. Trodden,
  Phys. Rev. D\textbf{78}, 023009 (2008)

\bibitem{ValiviitaAbdalla} J. Valiviita, E. Majerotto and R. Maartens,
JCAP 0807, 020 (2008); J.-H. He, B. Wang and E. Abdalla, Phys. Lett. B\textbf{671},
139 (2009) 

\bibitem{PSC} P.S. Corasaniti, Phys. Rev. D\textbf{78}, 083538 (2008)

\bibitem{Das} S. Das, P.S. Corasaniti and J. Khoury,
  Phys. Rev. D\textbf{73}, 083509 (2006)

\bibitem{kamion} M. C. Johnson and M. Kamionkowski,
Phys. Rev. D\textbf{78}, 063010 (2008)

\bibitem{bean08} R. Bean, E.E. Flanagan, I. Laszlo and M. Trodden,
Phys. Rev. D\textbf{78}, 123514 (2008)

\bibitem{mena09} M.B. Gavela, D. Hernandez, L. Lopez Honorez, O. Mena and
S. Rigolin, JCAP, 0907, 034 (2009)

\bibitem{lavacca09} G. La Vacca, J.R. Kristiansen, L.P.L. Colombo,
R. Mainini and S.A. Bonometto, JCAP, 0904, 007 (2009)

\bibitem{union} M. Kowalski et al., Astrophys. J. \textbf{686}, 749 (2008)

\bibitem{wmap5} J. Dunkley et al., Astrophys. J. Suppl. \textbf{180},
36 (2009)

\bibitem{sdss} M. Tegmark et al., Phys. Rev. D\textbf{74}, 123507
(2006)

\bibitem{zalda} U. Seljak and M. Zaldarriaga,
Astrophys. J. \textbf{469}, 437 (1996)


\end{thebibliography}

\IfFileExists{\jobname.bbl}{}
 {\typeout{}
  \typeout{******************************************}
  \typeout{** Please run "bibtex \jobname" to obtain}
  \typeout{** the bibliography and then re-run LaTeX}
  \typeout{** twice to fix the references!}
  \typeout{******************************************}
  \typeout{}
 }

\end{document}